\documentclass[aps,twocolumn,amsmath,amssymb,prb,superscriptaddress]{revtex4-1}
\usepackage{graphicx}
\usepackage{dcolumn}
\usepackage{bm}
\usepackage{color}	
\usepackage{ulem}	
\usepackage{braket}
\usepackage{mathrsfs}
\usepackage{float}
\usepackage{afterpage}
\usepackage[ppl]{mathcomp}

\begin{document}

\preprint{APS/PRB}

\title{Magnetic Phases of Frustrated Ferromagnetic Spin-Trimer System Gd$_{3}$Ru$_{4}$Al$_{12}$ 
With a Distorted Kagome Lattice Structure
}

\date{\today}

\author{S. Nakamura}
\thanks{coresponding authors}
\affiliation{Institute for Materials Research, Tohoku University, Katahira, Sendai 980-8577, Japan}
\affiliation{Center for Low Temperature Science, Tohoku University, Katahira, Sendai 980-8577, Japan}
  
\author{N. Kabeya}
\affiliation{Department of Physics, Tohoku University, Aramaki, Sendai 980-8578, Japan}
\affiliation{Center for Low Temperature Science, Tohoku University, Katahira, Sendai 980-8577, Japan}

\author{M. Kobayashi}
\affiliation{Department of Physics, Tohoku University, Aramaki, Sendai 980-8578, Japan}

\author{K. Araki}
\affiliation{Department of Applied Physics, National Defense Academy, Yokosuka 239-8686, Japan}

\author{K.  Katoh}
\affiliation{Department of Applied Physics, National Defense Academy, Yokosuka 239-8686, Japan}
 
\author{A. Ochiai}
\affiliation{Department of Physics, Tohoku University, Aramaki, Sendai 980-8578, Japan}
\affiliation{Center for Low Temperature Science, Tohoku University, Katahira, Sendai 980-8577, Japan}

\begin{abstract}
The magnetization and specific heat measurements have been performed on 
single-crystalline Gd$_{3}$Ru$_{4}$Al$_{12}$, wherein magnetic Gd--Al layers with a distorted 
Kagome lattice structure and non magnetic Ru--Al layers are stacked alternately along 
the $c$ axis.
A recent investigation has indicated that the distorted Kagome lattice structure of Gd--Al 
layers effectively translates into an antiferromagnetic triangular lattice in association 
with ferromagnetic spin trimerization at low temperatures.
We investigate the successive phase transitions and peculiar features of 
magnetic phases on this effective triangular lattice of spin trimers.
This spin system is found to be a $XY$ like Heisenberg model.
The magnetic phase diagrams indicate the existence of frustration
and ${\bm Z_{2}}$ degeneracy. The magnetization and specific heat imply the
successive phase transitions with partial disorder and a T-shaped spin structure 
in the ground state. 
\end{abstract}


\maketitle

\section{Introduction}

Metallic $4f$ frustrated spin systems often exhibit peculiar features at low temperatures.
Ternary intermetallic compounds RE$_{3}$Ru$_{4}$Al$_{12}$ (RE: rare earth) crystallize
in a hexagonal structure of Gd$_{3}$Ru$_{4}$Al$_{12}$-type, 
which belongs to the space group $P6_3/ mmc$
\cite{Niermann2002}.
In this crystal, magnetic RE-Al layers and 
non-magnetic Ru-Al layers stack alternately along the $c$ axis [Figs. 1 (a) and (b)]
\cite{VESTA}.
As shown in Fig. 1 (c), the RE ions form a distorted kagome lattice or a breathing kogome lattice
composed of two different sized regular triangles and unequal sided hexagons.
RE$_{3}$Ru$_{4}$Al$_{12}$ has been investigated intensively in recent years because of
the various phenomena it shows at low temperatures.
La$_{3}$Ru$_{4}$Al$_{12}$ is Pauli paramagnetic (PM) and Pr$_{3}$Ru$_{4}$Al$_{12}$ and 
Nd$_{3}$Ru$_{4}$Al$_{12}$ are ferromagnetic (FM)
\cite{Ge2012,Ge2014,Troc2012,Gorbunov2016}.
Ce$_{3}$Ru$_{4}$Al$_{12}$ is thought to be a valence fluctuation system 
\cite{Niermann2002}.
When the RE sites are replaced by heavy RE ions, RE$_{3}$Ru$_{4}$Al$_{12}$
shows antiferromagnetic (AFM) properties.
Yb$_{3}$Ru$_{4}$Al$_{12}$ is an $XY$-antiferromagnet
with N\'{e}el order at $T_{\rm N}=1.5$ K
\cite{Nakamura2014,Nakamura2015}.
This compound is a heavy fermion system with enhanced Sommerfeld coefficients
$\gamma_{0}=120$ mJ/(K$^{2}$ Yb-mol. 
Dy$_{3}$Ru$_{4}$Al$_{12}$ is an AFM compound with $T_{\rm N}=7$ K, which has a noncollinear 
spin structure 
\cite{Gorbunov2014}.
Regardless of the long range AFM ordering,
this compound shows a large $\gamma_{0}$ value of about 500 mJ/(K$^{2}$ Dy-mol)
in the temperature range 7-20 K.
Gorbunov {\it et al.} attributed this large $\gamma_{0}$ value to spin fluctuations 
induced in the Ru 4$d$ electrons by the exchange field acting from Dy 4$f$ electrons
\cite{Gorbunov2014}.
Chanragiri {\it et al.} have found characteristics of spin glass like
dynamics in Dy$_{3}$Ru$_{4}$Al$_{12}$ in AFM phase which indicates a
complex ground state under the influence of geometrical frustration 
\cite{Chandragiri2016}.
 		\begin{figure}[t]
		\begin{center}
		\includegraphics[width=8.6cm]{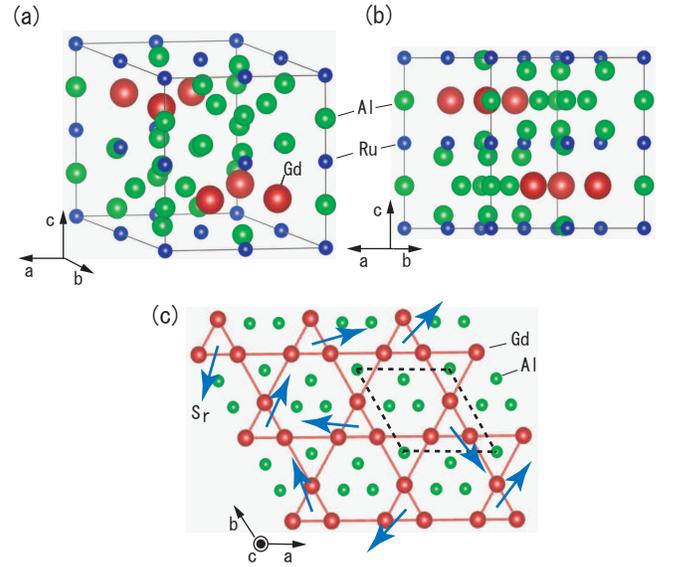}
		\end{center}
		\caption{(Color online) 
		(a) Structure of Gd$_{3}$Ru$_{4}$Al$_{12}$
		\cite{Niermann2002,VESTA}.
		The red (large), blue (small)
		and light green (middle) spheres denote Gd, Ru and Al ions, respectively. 
		(b)	Structure projected parallel to the $ab$ plane. 
		(c) A Gd--Al layer projected parallel to the $c$ axis.
		The red (larger) and 
		light green (smaller) spheres denote Gd and Al ions, respectively. 
		Bonds are drawn between the nearest neighbor and next nearest neighbor Gd ions. 
		The blue arrows indicate
		resultant spin $\bm S_{r}$ ($S_{r}=21/2$) on the FM trimers. 
		The broken rhombus indicates a unit cell. 
		}
		\label{Crystal}
		\end{figure}

In 2016, Chandragiri {\it et al.} reported the magnetic behavior of 
poly-crystalline Gd$_{3}$Ru$_{4}$Al$_{12}$,
whose magnetic susceptibility follows
the Curie--Weiss law above 200 K and 
whose Curie--Weiss temperature ($\theta_{p}$) has been estimated to be +80 K
\cite{Chandragiri2016_2}.
The magnetic susceptibility begins to increase rapidly with temperature decreasing
below 50 K, which implies the development of a FM correlation between the spins. 
However, it exhibits a sharp peak at 18.5 K, indicating AFM order.
The magnetic specific heat exhibits a broad maximum around 50 K, suggesting a glassy ground state.
On the other hand, the magnetic susceptibility exhibits a very small difference under 
zero field cool (ZFC) and field cool (FC) conditions.
The behavior of the magnetic susceptibility under magnetic fields is mimics
that expected for the Griffiths phase
\cite{Griffiths1969}.
Very recently, Nakamura {\it et al.} investigated the low-temperature magnetic and thermodynamic
properties of single-crystalline Gd$_{3}$Ru$_{4}$Al$_{12}$
\cite{Nakamura2018}.
They proposed that ferromagnetic (FM) spin trimers are formed
on small Gd-triangles at low temperatures, and that the distorted Kagome lattice of 
Gd$_{3}$Ru$_{4}$Al$_{12}$ effectively transforms into an antiferromagnetic
triangular lattice (AFMTL) at low temperatures. 
The blue arrows $\bm{S_{r}}$ in Fig.~\ref{Crystal}
denote the resultant spin ($S_{r}=21/2$) formed by the Ruderman--Kittel--Kasuya--Yosida 
(RKKY) interaction on the trimers.
These $\bm{S_{r}}$'s begin to be formed around 150 K and are completed below 70 K.
The binding energy is thought to be 184 K per Gd ion.
On further decreasing temperature, Gd$_{3}$Ru$_{4}$Al$_{12}$ exhibit successive AFM phase transition at
$T_{2}=18.6$ K and $T_{1}=17.5$ K.
The magnetic entropy at $T_{2}=18.6$ K is
only 40\% of $R {\rm ln}8$, indicating spin frustration.
Because binding energy is much higher than that at these transition temperatures,
the FM trimers are probably stable even in the ordered phases.

The ground state and magnetic phase diagrams of two-dimensional (2D) AFMTL's and 
three-dimensional (3D), or layered AFMTL's of Heisenberg models and related models 
(Heisenberg-Ising and Heisenberg-$XY$ models)
have been extensively investigated for long years from the view point of
geometrical frustration
\cite{Kawamura1998Review}. 
On the other hand, the oscillatory features of the RKKY interaction lead to the
frustration arising from the competition between the near and far-neighbor
interactions, which induce the spin glass in random system and spiral magnets
in periodic systems
\cite{Kawamura1998Review}. 
In the case of Gd$_{3}$Ru$_{4}$Al$_{12}$, the long range and oscillatory feature of the RKKY interaction 
also induces a geometrical frustration in association with the formation of FM trimers
at low temperatures
\cite{Nakamura2018}.
The present paper addresses the spin structures in the ordered phases and magnetic phase 
diagrams of the layered frustrated spin trimer system Gd$_{3}$Ru$_{4}$Al$_{12}$ wherein 
the geometrical and the interaction-compete-type frustrations coexist. The $\bm{S_{r}}$
system in Gd$_{3}$Ru$_{4}$Al$_{12}$ is regarded as an AFMTL lattice of the Heisenberg model 
with a certain degree strong anisotropy and interlayer interactions at low temperatures.
The long reaching range of the RKKY interaction may lead to some clear appearances of 
the geometrical frustration regardless of a slightly complicated geometrical structure of
the distorted kagome lattice.

\section{Typical magnetic phase diagrams with weak anisotropy}

The Hamiltonian of 2D Heisenberg model with weak anisotropy on AFMTL's under the 
field is written as,
	\begin{align}
		\mathscr{H}=J\sum_{i,j}{\bm S_{i}}{\bm S_{j}}-D\sum_{i}(S_{i}^{z})^{2}
		+g_{s}\mu_{\rm B}H_{z}\sum_{i}S_{i}^{z}.
		\label{Hamiltonian1}
	\end{align}
Here, the first term on the right side denotes the exchange interaction, the second term
denotes the local anisotropy at $i$ site, and the last term denotes the Zeeman energy. 
When $D$ is negative, the spin system is $XY$ like (easy plane type anisotropy), and
when $D$ is positive, the spin system is Ising like (easy axis type anisotropy).
Several theoretical investigations of frustrated AFMTL or layered AFMTL with anisotropy 
predict two successive phase transitions when $D>0$ at zero field
\cite{MiyashitaKawamura1985,Miyashita1986,Kawamura1998Review}.
In this case, the spin component along the easy axis and the other spin components
are ordered at distinct temperatures. In the case where the anisotropy is relatively strong,
three successive phase transitions are expected
\cite{Melchy2009}.
On the other hand, when $D<0$, only single-phase transition is expected at zero field
\cite{MiyashitaKawamura1985,Miyashita1986,Kawamura1998Review}.

  		\begin{figure}[h]
		\begin{center}
		\includegraphics[width=8cm]{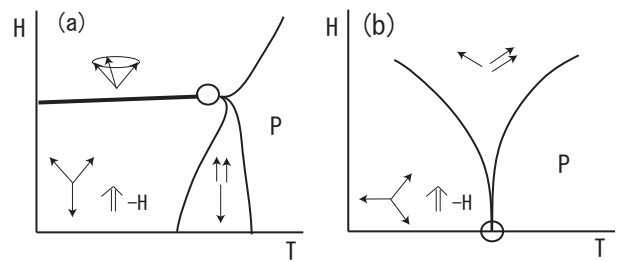}
		\end{center}
		\caption{Schematic magnetic phase diagrams of frustrated layered AFMTL with weak anisotropic
		interaction
		\cite{Kawamura1990,Kawamura1998Review}.
		(a) Easy axis type. The bold line indicates first-order transition and the open circle denotes
		a tetracritical point. Fields are directed along the easy axis. (b) Easy plane type. The
		open circle denotes a tetracritical point. Fields are directed parallel to the easy plane. 
		The minus sign of ${\bm H}$ denotes that the magnetic moments are in the opposite directions 
		of the spins.
		}
		\label{Theory}
		\end{figure}

The Hamiltonian of the layered Heisenberg model with weak anisotropy on AFMTL's
under the field is written as
	\begin{align}
		\mathscr{H}&=J\sum_{i,j}{\bm S_{i}}{\bm S_{j}}+J'\sum_{i,j}{\bm S_{i}}{\bm S_{j}}
						-D\sum_{i}(S_{i}^{z})^{2}\notag \\
						&\quad +g_{s}\mu_{\rm B}H_{z}\sum_{i}S_{i}^{z}.
		\label{Hamiltonian2}
	\end{align}
Here, the first term on the right side indicates intralayer exchange interaction
and the second term indicates interlayer exchange interaction.
When the anisotropy is the easy axis type ($D>0$), two successive phase transitions
are expected at zero field, similar to the 2D lattice
\cite{Kawamura1990,Kawamura1998Review}. 
We illustrate schematic phase diagrams in Fig.~\ref{Theory} according to these previous studies.
In the IMT phase shown in Fig.~\ref{Theory} (a), only longitudinal
spin component is ordered. The ground state is the noncolinear spin structure.
This state translates to the umbrella structure at high fields in association with the first order
phase transition when the field is applied along the easy axis. A tetracritical
point is predicted at the high temperature end of the first order boundary.
When $D<0$, only single-phase transition is expected
at zero field, similar to 2D system, and this 
transition point becomes a tetracritical point due to $\bm{Z_{2}}$ degeneracy.
CsNiCl$_{3}$ is known for a substance that shows a phase diagram such as that in 
Fig.~\ref{Theory} (a)
\cite{Clark1972,Poirier1990,Backmann1993,Kadowaki1987,Maegawa1988}
and CsMnBr$_{3}$ is
for a substance that shows a diagram such as that in Fig.~\ref{Theory} (b)
\cite{Gaulin1989}.

\section{Sample preparation and experimental method}

We melted 3N-Gd, 3N-Ru, and 5N-Al in a tetra-arc furnace and pulled 
a single-crystal ingot.
Considering evaporation loss, the initial weight of Al
was increased by 1--2\% in comparison to the stoichiometric amount.
The obtained ingot was about 2--3 cm in length and 3 mm in diameter. 
We determined the crystal structure of the ingot by X-ray diffraction
with crushed powder samples.
The diffraction pattern was consistent with that of a previous report
\cite{Niermann2002}. 
The lattice constants of Gd$_{3}$Ru$_{4}$Al$_{12}$
were obtained as 0.8778 nm for the $a$ axis and 0.9472 nm for the $c$ axis.
The length of the side of the small regular triangle was 0.3698 nm and that of the
large regular triangle was 0.5079 nm.
We cut three crystal samples from the ingot, 
one for magnetization measurements of 29.55 mg and the others for specific heat 
measurements of 7.76 mg and 13.99 mg. All samples are the same as those used in the
previous investigation
\cite{Nakamura2018}.
The specific heat measurements of the specific heat were performed by a thermal relaxation method using
a commercial instrument (PPMS-9, Quantum Design Inc.) above 2 K 
and a quasi-adiabathic method with a hand-made instrument below 2 K.
The magnetization
was measured using two superconducting quantum interference device magnetometers
(MPMS, Quantum Design Inc.).

\section{Experimental Results}

\subsection{The magnetic phase transition with changing temperature}

\begin{figure}[b]
		\begin{center}
		\includegraphics[width=8cm]{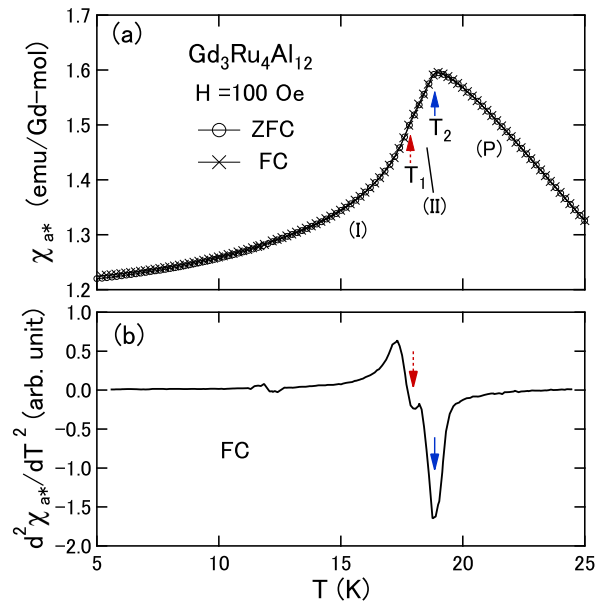}
		\end{center}
		\caption{(Color online) (a) Temperature dependence of magnetic susceptibility 
		$\chi_{a*}$.  
		The open circles and crosses denote $\chi$ measured in ZFC
		(5$\rightarrow$25 K)
		and FC (25$\rightarrow$5 K) processes under a field of 100 Oe.
		The broken red and solid blue upward arrows indicate phase transition
		temperatures. The strength of the fields contains several Oe errors.
		(b) The second derivative of $\chi_{a*}$ (FC) in relation with temperature.
		The broken red and solid blue downward arrows correspond to inflection points in 
		$\chi_{a*}$.
		}
		\label{Sus_ZFCFC}
		\end{figure}

The temperature dependence of magnetic susceptibility $\chi_{a*}$ (${\bm H}\parallel a*$)
of Gd$_{3}$Ru$_{4}$Al$_{12}$ is shown in Fig.~\ref{Sus_ZFCFC} (a).
The open circles and crosses denote the ZFC
and FC processes under a field of 100 Oe, respectively.
Both $\chi_{a*}$ exhibit very small differences
between the ZFC and FC processes.
Because the applied magnetic field is weak, these results include
few percent error in the absolute values.
The upward arrows in Fig.~\ref{Sus_ZFCFC} (a) indicate phase transition points.
Figure~\ref{Sus_ZFCFC} (b) shows the second derivatives of $\chi_{a*}$ in 
relation to temperature. 
We identify the inflexion points in $\chi_{a*}$ as the transition points.
The weak anomalies shown in Fig.~\ref{Sus_ZFCFC} (b) at 12 K arise from thermocouple 
conversion in MPMS and are not essential.
In the present paper, we refer to the lower and
higher transition temperatures as $T_{1}$ and $T_{2}$,
and low temperature phase and intermediate temperature (IMT) phase 
as phase I and phase II, respectively, in accordance with the previous report
\cite{Nakamura2018}.

\begin{figure*}[t]
		\begin{center}
			\includegraphics[width=15cm]{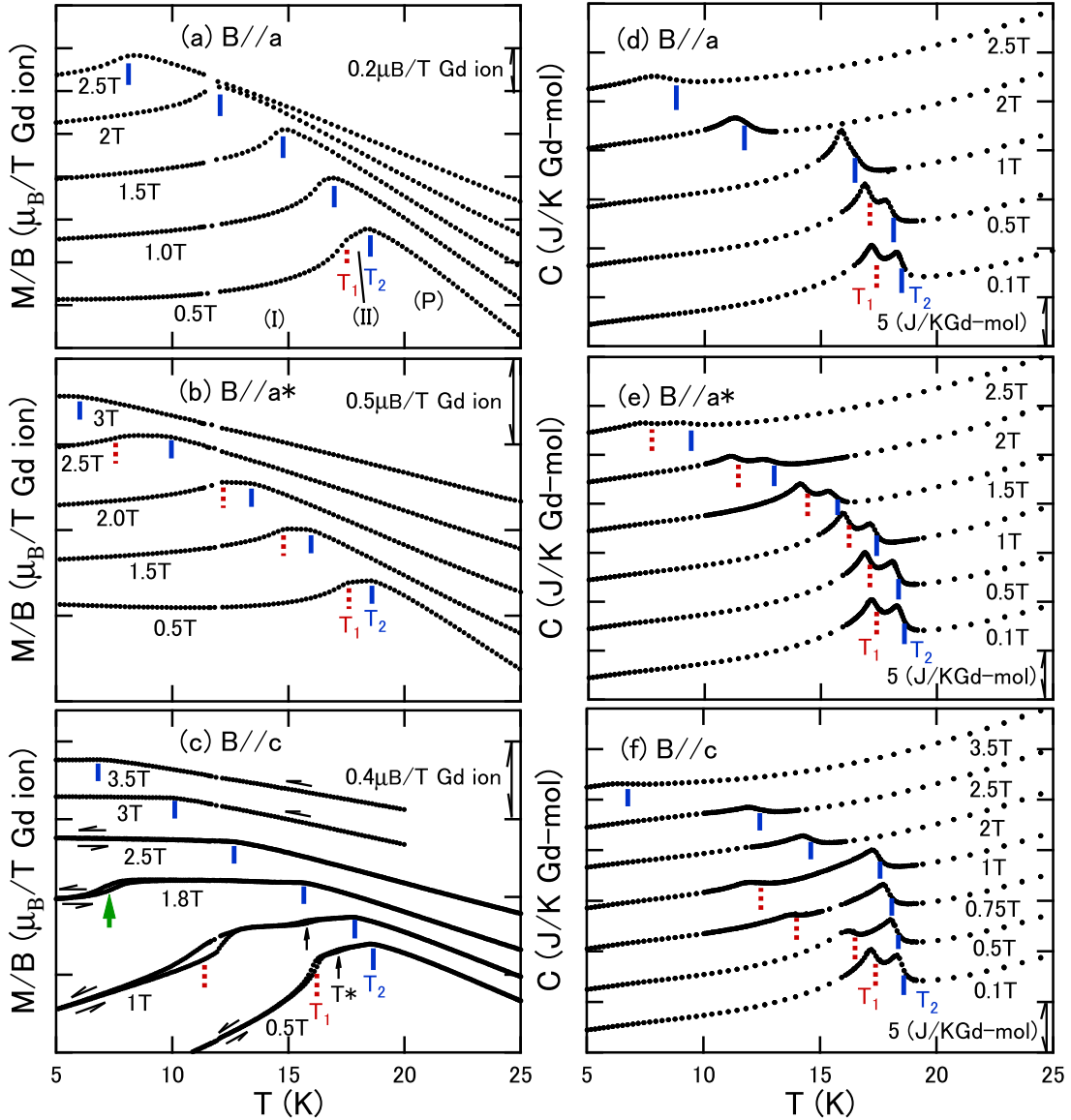}
		\end{center}
		\caption{(Color online) Phase transition points of Gd$_{3}$Ru$_{4}$Al$_{12}$
		observed (a)--(c) in the temperature dependence of magnetization and 
		(d)--(f) in the specific heat.
		The fields are directed along the (a), (d) $a$ axis; (b), (d) $a^{*}$ axis; (e) and (f) 
		$c$ axis. 
		The red dotted lines and blue solid lines commonly indicate $T_{1}$ and $T_{2}$, respectively.
		The bold green arrow in panel (c) indicates the phase II/phase III transition
		point at 1.8 T.
		The origin of each set of data is shifted for ease of viewing. $T^{*}$ in (c) panel
		denotes the weak anomalies that may not be phase transition points (see text).
		}
		\label{MTandCT}
		\end{figure*}

Selected temperature dependence of the magnetic susceptibility $M/B$ and specific heat 
at several fields is presented in Fig.~\ref{MTandCT},
where $T_{1}$ and $T_{2}$ are commonly indicated by the red dotted lines and
blue solid lines, respectively.
Figure~\ref{MTandCT} (a) shows $M/B$ under fields directed along the $a$ axis. 
The measurements were performed with FC processes and
we identified the reflection points of $M/B$ as the phase transition points.
When fields are applied along the $a$ axis, the IMT phase II
only appears in the low field range. Gd$_{3}$Ru$_{4}$Al$_{12}$ directly translates
from the PM phase into phase I in the high field range.

Magnetic susceptibility $M/B$
under several fields directed along the $a^{*}$ axis are presented in Fig.~\ref{MTandCT} (b). 
The measurements were performed FC processes.
When fields are applied along the $a^{*}$ axis, phase II
appears even in the high field range.
As evident in Fig.~\ref{MTandCT} (b), one of the characteristic features of the IMT phase II
is the weak temperature dependence in $M/B(T)$. In other words, $M/B$ behaves
like a transverse susceptibility in phase II.

In Fig.~\ref{MTandCT} (c), magnetic susceptibility $M/B$ under fields
directed along the $c$ axis are presented.
The measurements were performed with FC and field heat (FH) processes in succession at 0.5, 1,
1.8 and 2.5 T, and with FC process at 3 and 3.5 T.
When the field is directed along the $c$ axis, $M/B(T)$ shows hysteresis loops
at $T_{1}$ in the range $0.3\le B \le 2$ T. We identified  the inflection points in 
$M/B$ as $T_{2}$ and centers of the hysteresis loops as $T_{1}$.
The bold green upward arrow denotes the phase II/phase III transition points at 1.8 T.
We have found an additional phase III in the intermediate fields for ${\bm B} \parallel c$.
As indicated in Fig.~\ref{MTandCT} (c) by black upward arrows and symbol
$T^{*}$, small anomalies are observed between $T_{1}$ and $T_{2}$ in the
field range $0.3\le B\le 1.5$ T. 
However, we could not observe
any anomaly in the specific heat at $T^{*}$ as mentioned later.
Probably, the anomalies at $T^{*}$ in $M/B$ does not indicate phase transition.
As shown in Fig.~\ref{MTandCT} (c),
the IMT phase II appears over the wide temperature ranges in the intermediate field range. 
Attention should be paid to the temperature dependence of $M/B$ in phase II.
When fields are weak, $M/B$ shows some temperature dependence in phase II, but when the field 
becomes slightly strong, $M/B$ is almost temperature-independent in this phase.
Apparently, $M/B$ is a transverse susceptibility in phase II at slightly strong fields.
In phase I, $M/B$ shows larger temperature dependence. Apparently, the component of
the longitudinal magnetic susceptibility exists in phase I.

The specific heat $C$ at several fields under the fields directed along the $a$ axis are presented in Fig.~\ref{MTandCT} (d).
Corresponding to the successive phase transitions at $T_{1}$ and $T_{2}$,
clear $\lambda$-shaped peaks are observed in the specific heat at low fields.
In the present study, we identified the phase transition points as the middle points
on the right-side slopes of the peaks.
The two peaks shown at low fields change into a single peak at high fields.
This behavior of the transition points is consistent with that observed in the $M/B$ shown in
Fig.~\ref{MTandCT} (a).

In Fig.~\ref{MTandCT} (e), specific heat $C$ at several fields under the fields 
directed along the $a^{*}$ axis are shown.
Corresponding to the successive phase transitions at $T_{1}$ and $T_{2}$,
clear $\lambda$-shaped peaks are observed as well.
The IMT phase II is observed even in high fields similar to the case of observation of the 
magnetic susceptibility presented in Fig.~\ref{MTandCT} (b).

Specific heat $C$ at several fields under the fields 
directed along the $c$ axis are presented in Fig.~\ref{MTandCT} (f).
Clear $\lambda$-shaped peaks are observed at $T_{1}$ and $T_{2}$.
The IMT phase II occupies a wide temperature range at intermediate field range.
We could not find any indication of phase transition at $T^{*}$ in the specific heat.
Probably, the anomalies at $T^{*}$ are so not indicate phase transitions.
They may indicate certain domain motion in Phase II.

\subsection{The magnetic phase transitions with changing field}

		\begin{figure}[h]
		\begin{center}
			\includegraphics[width=8.6cm]{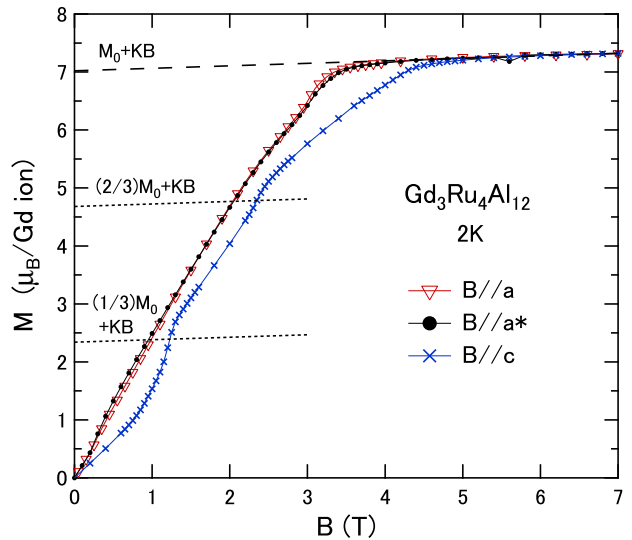}
		\end{center}
		\caption{(Color online) Magnetization curves of Gd$_{3}$Ru$_{4}$Al$_{12}$
		at 2 K in a field increasing process.
		The fields are directed along the $a$, $a^{*}$ and $c$ axes. 
		The broken line is a fit to the formula $M(B)=M_{0}+KB$
		in the range $5.6<B<7$ T for ${\bm B} \parallel c$.
		Here, $M_{0}$ is a constant independent of the field and $K$ is a proportion
		constant.
		The doted lines are guides for eye denoting the functions 
		$M(B)=FM_{0}+KB$
		$(F=1/3$, 2/3). 
		}
		\label{MH_2K}
		\end{figure}

		\begin{figure}[h]
		\begin{center}
			\includegraphics[width=8.6cm]{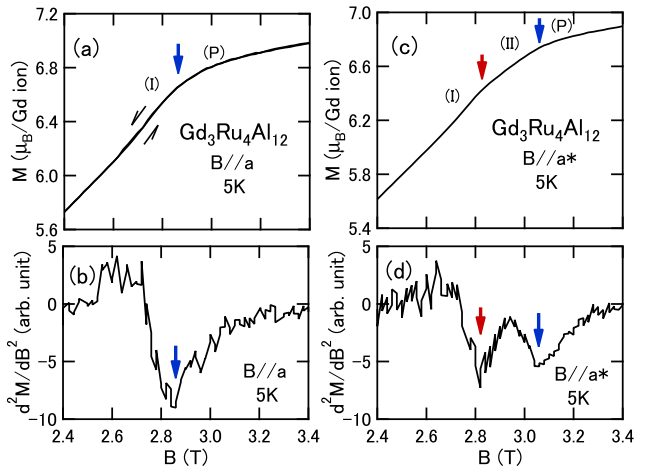}
		\end{center}
		\caption{(Color online) (a) Data of $M (B)$ 
		of Gd$_{3}$Ru$_{4}$Al$_{12}$ for ${\bm B} \parallel a$ at 5 K.
		(b) The second derivative of $M$ in panel (a) with the elevating field process. 
		(c) Data of $M (B)$ Gd$_{3}$Ru$_{4}$Al$_{12}$ for ${\bm B} \parallel a*$.
		(d) The second derivative of $M$ in panel (c).
		The arrows are indications of phase transition points.
		}
		\label{MH5K_A}
		\end{figure}

The field dependence of magnetization $M$ of Gd$_{3}$Ru$_{4}$Al$_{12}$ at 2 K is displayed 
in Fig.~\ref{MH_2K}.
Overall, the magnetic anisotropy is clearly seen, {\it i.e.}
$ab$ is an easy plane of magnetization and $c$ is a difficult axis of magnetization.
The anisotropy in the $ab$ plane is very small.
The additional phase III appears in the intermediate field range $1.25 <B <2.4$ T
when the field is applied along the $c$ axis at 2 K.
Regardless of the difference in field direction, $M$  
shows a tendency to increase approximately linearly with 
magnetic field in the high field range. 
We assume that $M$ at high fields can be described using $M(B)=M_{0}+KB$.
Here, $M_{0}$ is a constant that does not depend on the field and 
$K$ is a proportion constant.
The broken line in Fig.~\ref{MH_2K} is a fit to the data for ${\bm B} \parallel c$ 
in the range $5.6<B<7$ T. 
The magnetization $M_{0}=7.02$ $\mu_{\rm B}$ obtained for ${\bm B}\parallel c$ 
agrees with that expected for Gd$^{3+}$ ($S=7/2$).
The proportion constant $K$ is estimated to be 4.3$\times 10^{-2}$ $\mu_{\rm B}$T$^{-1}$
(2.4$\times10^{-2}$ emu).
If we assume that $K$ arises from Pauli paramagnetism from Ru 4$d$ electrons,
it is three orders larger than that for usual transition metals
\cite{Kriessman1954}.
However, this is not the heavy fermion behavior. As we mention later, the low temperature
specific heat of Gd$_{3}$Ru$_{4}$Al$_{12}$ is not $T$-linear in the very low temperature range.
To determine the accurate magnetization processes as field functions, more precise and wide range 
measurements in the high field range are needed.
As shown in Fig.~\ref{MH_2K}, two spin-flopping-like anomalies appear 
in $M$ for ${\bm B} \parallel c$ axis at around 1.25 and 2.4 T. 
The dotted lines in Fig.~\ref{MH_2K}
are $M$ calculated from the formula $M(B)=FM_{0}+B$ ($F=1/3$, 2/3).
Apparently, the spin-flopping-like anomalies appear at the points where
the magnetization of Gd ions is approximately equal to $(1/3)M_{0}$ and $(2/3)M_{0}$.

		\begin{figure}[t]
		\begin{center}
			\includegraphics[width=8.6cm]{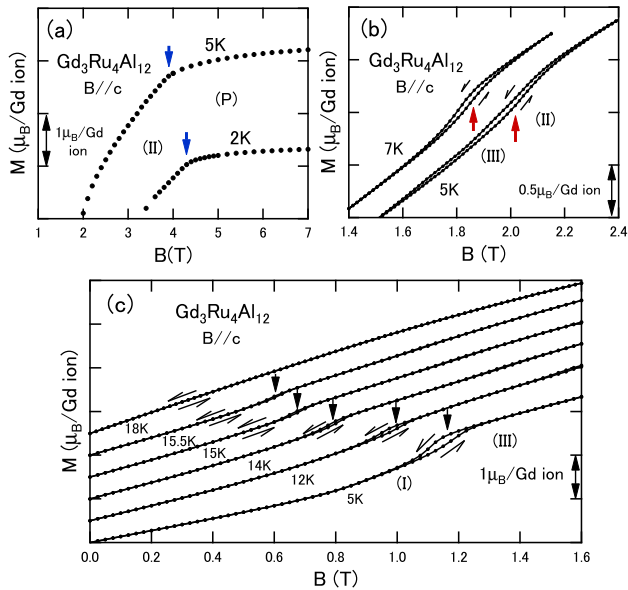}
		\end{center}
		\caption{(Color online) Data of $M (B)$ 
		of Gd$_{3}$Ru$_{4}$Al$_{12}$ for $B \parallel c$ at several temperatures.
		(a) $M (B)$ in the high field range, (b) in the intermediate field range, and
		(c) in the low field range. The arrows indicate the phase transition points.
		}
		\label{MH5K_C}
		\end{figure}

Figure~\ref{MH5K_A} (a) presents the $M (B)$ curve for ${\bm B} \parallel a$ at 5 K.
The blue downward arrow indicates the phase I/PM phase transition point at 2.86 T.
Here, we regard the reflection point as the phase transition point.
Figure~\ref{MH5K_A} (b) shows the second derivative of $M$ in panel (a) 
with elevating field process. The minimum point in this figure corresponds to the reflection 
point. When the field is directed along the $a$ axis, Gd$_{3}$Ru$_{4}$Al$_{12}$
translates from phase I to PM phase directly.

Fig.~\ref{MH5K_A} (c) shows the $M (B)$ curve for $B \parallel a^{*}$ at 5 K, 
and Fig.~\ref{MH5K_A} (d) shows the second derivative of $M$ in panel (c).
The red arrows at a lower field side and the blue arrows at a higher field side
indicate the phase I/phase II transition point at 2.82 T and the phase II/PM phase
transition point at 3.06 T.
The minimum points shown in the second derivative of $M$ shown in Fig.~\ref{MH5K_A} (d)
correspond to these transition points, respectively.
When the fields are directed along the $a^{*}$ axis, phase II appears in the intermediate
field range even at low temperatures.

		\begin{figure*}[t]
		\begin{center}
			\includegraphics[width=15cm]{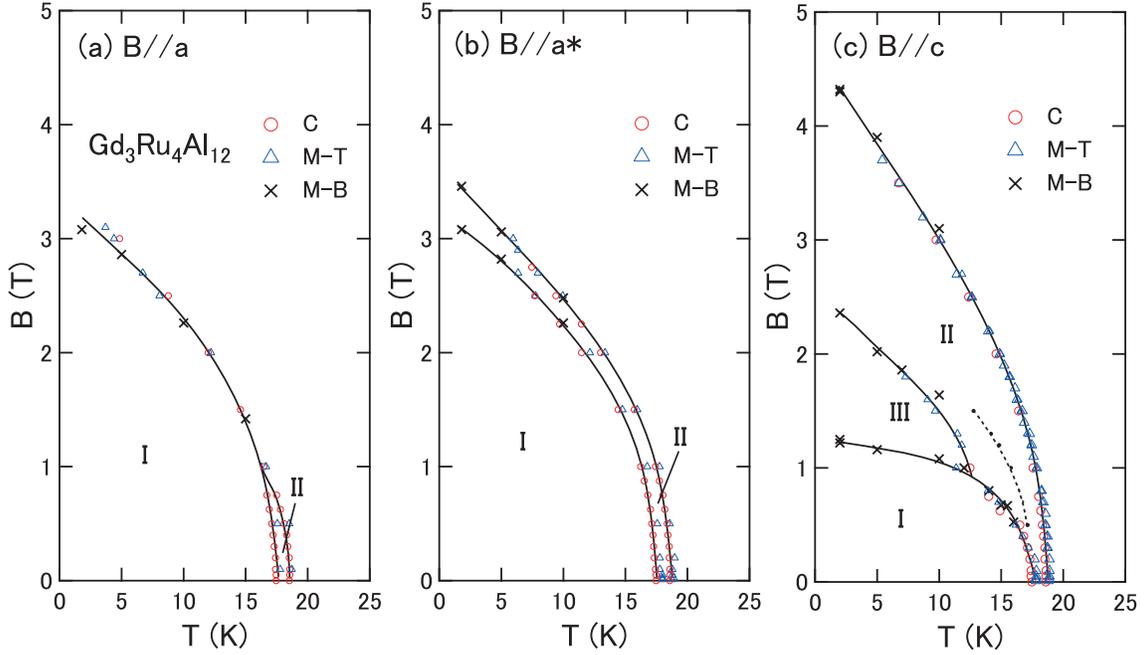}
		\end{center}
		\caption{(Color online) Magnetic phase diagrams of Gd$_{3}$Ru$_{4}$Al$_{12}$ for
		(a) ${\bm B} \parallel a$, (b)  ${\bm B} \parallel a^{*}$, and (c) ${\bm B} \parallel c$ 
		axes. The red circles, blue triangles, and black crosses indicate the phase transition 
		points determined from specific heat, and $M/B(T)$ and $M/B(B)$ measurements.
		Data for $C$ at zero field are taken from the reference
		\cite{Nakamura2018}. 
		The dotted line in the right panel 
		corresponds to $T^{*}$ shown in Fig. 4 (c). This line may not be
		a phase boundary.
		}
		\label{PhaseD}
		\end{figure*}

Figure~\ref{MH5K_C} displays the magnetization curves under the fields
directed along the $c$ axis. The solid downward blue arrows in Fig.~\ref{MH5K_C} (a) 
indicate the phase II/PM phase transition that occurs at high fields. 
The magnetization curves in the intermediate field range shows small
hysteresis loops, as shown in Fig.~\ref{MH5K_C} (b).
The upward red arrows indicate phase III/phase II transitions and the hysteresis loops
imply that this transition is of first order. 
Similar small hysteresis loops are shown in the low field range, as shown
in Fig.~\ref{MH5K_C} (c). The black arrows indicate the phase I/phase III transitions.
The hysteresis loops imply that this phase transition is of first order as well.
The additional phase III is observed in the intermediate field range when
fields are directed along the $c$ axis, which is the hard axis of magnetization. 
This implies that phase III is induced with spin flopping.
It is probable that Gd$_{3}$Ru$_{4}$Al$_{12}$ undergoes two successive spin flopping,
when fields are applied along the $c$ axis.

\subsection{Magnetic phase diagrams}

Analyzing the results of measurements of magnetic susceptibility, magnetization, and specific heat,
we determined the magnetic phase diagrams of Gd$_{3}$Ru$_{4}$Al$_{12}$, as depicted
in Fig.~\ref{PhaseD}. 
The whole view of the magnetic phase diagrams presented in Fig.~\ref{PhaseD} 
is unexpectedly anisotropic for Gd compounds.
They look different from the phase diagrams of non-frustrated AFM spin systems. 
The existence of the IMT phase II implies the existence of geometrical frustration. 
However, there are several features different from the phase diagrams of the typical frustrated 
AFMTL's with weak anisotropy and weak interlayer interactions shown in Fig.~\ref{Theory}
in terms of the particulars. Let us take a look at the details.
Two successive AFM phase transitions have been observed at zero field.
This feature is different from that of the phase diagram in Fig.~\ref{Theory} (b).
Two double critical points, or N\'{e}el points, exist at zero field
instead of the single tetracritical point.
For $B \parallel a$, the AFM phase I occupies the low-$T$ and low-$B$ regions.
Between phase I and the PM phase, phase II occupies a strip region at low fields.
At a glance, this strip region appears similar to that shown in Fig.~\ref{Theory} (a).
However, the first-phase transition line shown in Fig.~\ref{Theory} (a) is not observed
in Fig.~\ref{PhaseD} (a).  
In addition, as shown in Fig.~\ref{PhaseD} (a), phase I directly contacts the PM phase 
with a boundary in the high field range. On the other hand, there is a high field phase with 
umbrella spin structure in Fig.~\ref{Theory} (a).
For $B \parallel a^{*}$, the boundaries of phase I/phase II
and phase II/PM phase display double lines that do not cross and
show the difference from non-frustrated AFM spin systems.
Probably, these double lines are clear appearance of frustration. 
When the field is applied along the $c$ axis, as shown in Fig.~\ref{PhaseD} (c),
phase III appears between phase I and phase II in the intermediate field range
and phase II relatively occupies a wide region in the diagram.
As mentioned before, the magnetization shows hysteresis loops at the
phase I/phase III and phase III/phase II transition points, and therefore, 
both these transitions are of first order.
The dotted line in Fig.~\ref{PhaseD} (c) corresponds to weak anomalies at $T*$ shown in 
Fig.~\ref{MTandCT} (c). This line may not be the phase boundary and may correspond to
certain domain motion.

The phase diagrams in Fig.~\ref{PhaseD} appear as if they are
a superposition of two independent non-frustrated AFM spin systems with different anisotropies,
at a glance.
One is the spin system that has easy plane (the $ab$ plane) type and the other is that
having easy axis (the $c$ axis) type.
The easy plane-type spin system exhibits a simple single-phase boundary
and the easy axis-type spin system shows spin flopping when fields are
applied along the $c$ axis, as shown in Fig.~\ref{PhaseD} (c) and Fig.~\ref{MH_2K}.
However, as evident from Figs.~\ref{PhaseD}, there is a feature
we cannot understand as the superposition of two independent spin systems.
Noted that phase I appears as a lower-temperature 
phase of phase II but phase II does not appear as a lower-temperature phase of phase I. 
This implies that these phases do not appear independently.
Overall, the magnetic phase diagrams of Gd$_{3}$Ru$_{4}$Al$_{12}$ indicate
the existence of frustration, but present several distinct appearances from
those of the typical Heisenberg model with weak anisotropy and weak interlayer interactions
on layered AFMTL's.

\section{Discussion}

\subsection{Single trimer magnetic anisotropy}

\begin{figure}[t]
		\begin{center}
			\includegraphics[width=8cm]{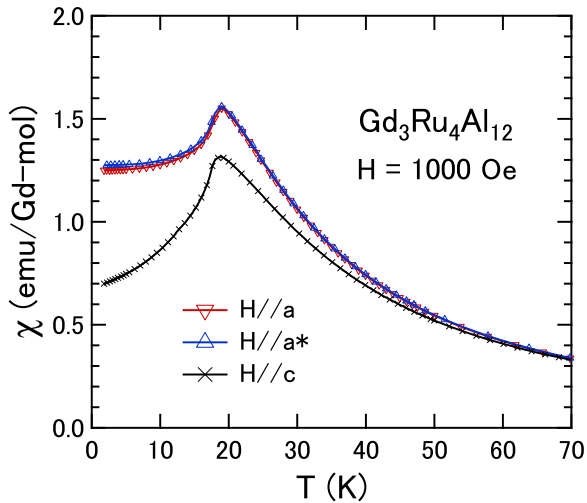}
		\end{center}
		\caption{(Color online) Temperature dependence of magnetic susceptibility 
		of Gd$_{3}$Ru$_{4}$Al$_{12}$. The applied field is 1000 Oe.
		The data are taken from the reference\cite{Nakamura2018}.	 		
		}
		\label{Sus_aniso}
		\end{figure}

\begin{figure}[b]
		\begin{center}
			\includegraphics[width=3.5cm]{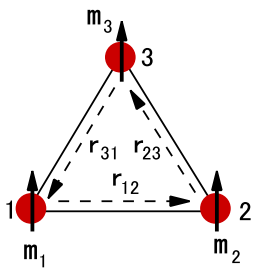}
		\end{center}
		\caption{(Color online) Magnetic moments $\bm m_{i}$ ($m_{i}=7\mu_{\rm B}$) on an FM trimer.
		Here, the subscripts $i=1,2,3$ denote the vertices of the triangle.
		The red spheres indicate Gd ions.
		The vector $\bm r_{ij}$ denotes the position of the vertex $j$ from vertex $i$.	 		
		}
		\label{MuonTrimer}
		\end{figure}

In Gd$_{3}$Ru$_{4}$Al$_{12}$, FM trimers ($S_{r}=21/2$) form the AFMTL at 
low temperatures
\cite{Nakamura2018}.
First, we discuss the single trimer anisotropy.
As shown in Fig.~\ref{Sus_aniso},
the magnetic anisotropy is observed even in the PM phase in the temperature range below 70 K, 
where ${\bm S_{r}}$ are completed
\cite{Nakamura2018}.
This suggests that magnetic anisotropy is induced by the formation of FM trimer.
One possible origin of anisotropy is electromagnetic interaction.
Figure~\ref{MuonTrimer} displays an FM trimer on which three magnetic moments 
$\bm m_{i}$ ($i=1,2,3; m_{i}=7\mu_{\rm B}$) are placed. Here, the subscripts $i=1,2,3$ indicate 
the number of vertices, and $\mu_{\rm B}=927.400\times10^{-26}$ JT$^{-1}$
is the Bohr magneton. The vector $\bm r_{ij}$ denotes the position of vertex $j$
from vertex $i$. The flux density $\bm B_{ij}$ at the vertex $j$ induced by $\bm m_{i}$ 
at the vertex $i$ is given by
		\begin{align}
			{\bm B_{ij}}=-\frac{\mu_{0}}{4\pi  r^{3}}\left[{\bm m_{i}}
					-\frac{3({\bm m_{i}}\cdot{\bm r_{ij}}){\bm r_{ij}}}{r^{2}}\right].\notag		
		\end{align}
When the FM trimer is formed at low temperatures, all three magnetic moments are written as 
${\bm m}$.
Therefore, the electromagnetic energy of the trimer is 
		\begin{align}
			E_{em}=\frac{ \mu_{0}}{4\pi r^{3}}\biggl[3{\bm m}^{2}
				-3\sum_{ij}\frac{({\bm m}\cdot{\bm r_{ij}})^{2}}{r^{2}}\biggr]	\notag
		\end{align}
at a unit of ${\rm J}$ per ${\bm S_{r}}$, where the suffix runs over $(ij=12,23,31)$. 
This energy becomes the lowest when ${\bm m}$ is directed in the $ab$ plane. 
The electromagnetic energy $E_{em}$ gives rise to the easy plane-type anisotropy,
and gives isotropy in the $ab$ plane. 
However, the amplitude of this energy is approximately
2.7 K per ${\bm S_{r}}$. This is too small to explain the anisotropy experimentally observed 
only for that, as mentioned later.

Another possible origin of the single trimer anisotropy is the generation of the orbital 
angular momentum of Gd$^{3+}$ ($4f^{7}, S=7/2$) ions. 
The 4$f$ electrons of Gd ions do not carry orbital angular momentum in general.
However, in the case of Gd$_{3}$Ru$_{4}$Al$_{12}$, Gd ions occupy the asymmetric site in the crystal. 
Therefore, the ions would feel odd parity CEF at each site,
which induces the mixing between the $4f$ and $3d$ electrons of the Gd ion, and
the Gd ions obtain some angular orbital momentum. This would result in single ion anisotropy.
In addition, the existence of orbital moments can lead to spatially anisotropic RKKY interactions
\cite{Timm2005},
which may induce single trimer anisotropy through a similar mechanism to the case of the 
above electromagnetic interaction, but detailed mechanism is unknown at present.
Probably, a combined effect of the anisotropy due to the odd parity CEF and the 
electromagnetic interaction is the origin of the single trimer anisotropy.

\begin{figure}[h]
		\begin{center}
			\includegraphics[width=8cm]{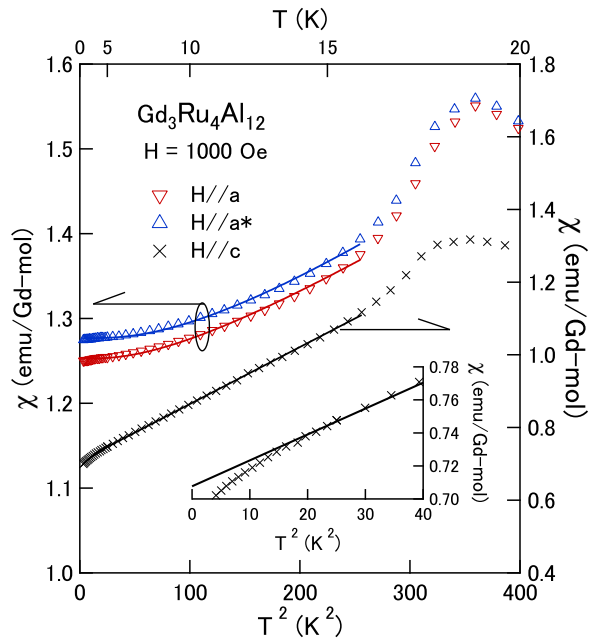}
		\end{center}
		\caption{(Color online) Temperature dependence of the magnetic susceptibility 
		of Gd$_{3}$Ru$_{4}$Al$_{12}$. 
		The data is taken from Fig.~\ref{Sus_aniso} and replotted on a $T^{2}$-scale. 
		The applied field is 1000 Oe.
		The solid curved lines are fits to Eq.~\ref{Gapless} and 
		the solid straight line is a fit to Eq.~\ref{Gap}.
		These lines also indicate the fitting regions. 
		The magnetic susceptibility for ${\bm H} \parallel c$ is
		presented in the inset on expanded scales. The fitting line is expanded to 
		the zero temperature in the inset.
		}
		\label{Sus_fit}
		\end{figure}

In any case, we need to determine the magnitude of the single trimer anisotropy in ordered 
phases experimentally.
The magnetic susceptibilities in Fig.~\ref{Sus_aniso}
at low temperatures are replotted in Fig.~\ref{Sus_fit} on expanded scales. In this figure,
magnetic susceptibilities are plotted as the function of $T^{2}$.
In phase I, AFM spin waves are expected to contribute to the magnetization at finite temperatures.
According to previous theories based on spin wave approximation, the contribution of the 
three-dimensionally propagating AFM spin waves can be expressed as
$M(T)-M(0)\propto T^{2}$ for isotropic systems
\cite{Kubo1952,QTS,Jaccarino}
and $M(T)-M(0)\propto T^{1.5}\exp{(-E_{g}/T)}$
for anisotropic systems
\cite{Jaccarino}
when the temperatures are sufficiently lower than N\'{e}el temperature.
Here, $M(0)$ is the magnetization at 0 K and   
$E_{g}$ is the energy gap in the AFM magnon dispersion,
	\begin{align}
		E_{g}=(nk_{\rm B})^{-1}\,\hbar\Omega=(nk_{\rm B})^{-1}\,
			\hbar\sqrt{\omega_{A}^{2}+2\omega_{ex}\omega_{A}},
		\label{Gapenergy}			
	\end{align}
at a unit of K per magnon.
Here, $n$ is the mole number of propagation medium ${\bm S_{r}}$'s, $\Omega$ the lowest precession 
frequency of magnons, $\hbar\omega_{A}=2\mu_{\rm B}B_{A}$  the crystal magnetic anisotropic energy 
on single trimer and $\hbar\omega_{ex}=2JSz$ the energy deduced by the exchange interactions 
from nearest neighbor ${\bm S_{r}}$'s of number $z$. The effective anisotropic flux density 
${\bm B_{A}}$ depends on the directions in general.
When the applied external flux density is sufficiently weak, $M$ can be replaced by $\chi$ as
	\begin{align}
		\chi(T)-\chi(0)&=C_{1}\,T^{2},
		\label{Gapless}\\
		\chi(T)-\chi(0)&=C_{2}\,T^{1.5}\exp{(-E_{g}/T)},
		\label{Gap}		
	\end{align}
where $C_{1}$ and $C_{2}$ are proportion constants.
When $E_{g}$ is large, the dispersion relation is given by,
		\begin{align}
			\hbar\omega(k)=\hbar\Omega+D_{a}k^{2},
		\label{Dispersion}		
	\end{align}
for small wave number $k$, where $D_{a}$ is proportion constant. 
The second term in the right side is similar to that for FM magnons. 
Thus, the numbers of the excited magnons at temperature $T$ is approximately 
in proportion to $T^{1.5}\exp{(-E_{g}/T)}$.

The solid blue and red curved lines in Fig.~\ref{Sus_fit} are fits to Eq.~\ref{Gap}.
The calculated data well reproduce the experimentally observed $\chi_{a}$ and $\chi_{a*}$.
The temperature dependence of these susceptibilities in the low temperature range can be 
understood as the contribution from three dimensionally propagate spin waves under ${\bm B_{A}}$.
The $E_{g}$ obtained are 24 K for $\chi_{a}$ and 29 K for $\chi_{a*}$,
being isotropic in the $ab$ plane.
On the other hand,
it can be seen that $\chi_{c}$ changes as a linear function of $T^{2}$ in the low 
temperature range.
The solid straight line in Fig.~\ref{Sus_fit} is a fit to Eq.~\ref{Gapless}.
The temperature dependence of $\chi_{c}$ is well explained by three dimensionally propagate 
spin wave contribution without ${\bm B_{A}}$ above 4.5 K.
It is inferred that single trimer anisotropy of Gd$_{3}$Ru$_{4}$Al$_{12}$ is easy plane type.
The ${\bm S_{r}}$'s which are parallel to the $ab$ plane 
feels relatively strong ${\bm B_{A}}$ along their directions,
and the others which are parallel to the $c$ axis only feel weak ${\bm B_{A}}$.
The observations of ${\bm B_{A}}$ indicate that the ${\bm S_{r}}$ system of Gd$_{3}$Ru$_{4}$Al$_{12}$ 
is an easy plane type, and the strength of the anisotropy is rather strong. 
This would have certain degree of characteristics of $XY$ model.

\subsection{Spin structure of the ground state}

\begin{figure}[h]
		\begin{center}
			\includegraphics[width=5cm]{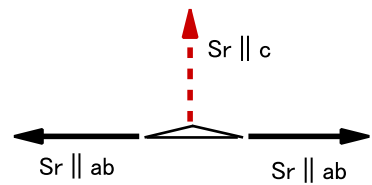}
		\end{center}
		\caption{(Color online) Spin structures of Gd$_{3}$Ru$_{4}$Al$_{12}$ in 
		phase I on the triangle of the trimers.
		The bold black arrows denote the resultant spin $\bm S_{r}$'s ($S_{r}=21/2)$
		directed in the $ab$ plane.
		The broken red arrows indicates the $\bm S_{r}$ directed perpendicular to the $c$ axis.		
		}
		\label{T_Structure}
		\end{figure}

When the anisotropy is weak, the ground state of AFMTL's are 
approximately the 120$^{\circ}$ structure. 
However, the actual anisotropy is not weak in Gd$_{3}$Ru$_{4}$Al$_{12}$. 
If the single trimer anisotropy is easy plane like, the basal plane of the 120$^{\circ}$
structure must be parallel to the $ab$ plane. 
In this case, it is difficult to explain the longitudinal component of magnetic susceptibility in
$\chi_{c}$ shown in Fig.~\ref{Sus_aniso}.
In addition, it is difficult to explain the the first order phase transition 
with spin flopping induced by the flux density ${\bm B} \parallel c$ shown in Fig.~\ref{PhaseD}
and Fig.~\ref{MH_2K}. The 120$^{\circ}$ structure would be change into the umbrella
structure in Fig.~\ref{Theory} (a) in the high field region without spin flopping.
Probably, we should consider some ground states of Gd$_{3}$Ru$_{4}$Al$_{12}$
being different from the 120$^{\circ}$ structure. 
Instead of the structure, let us examine the T-shaped
structure shown in Fig.~\ref{T_Structure}. In this figure, three ${\bm S_{r}}$'s are on the
vertexes of the triangle. A pair of ${\bm S_{r}}$'s depicted by solid black arrows in opposite
directions are directed parallel to the $ab$ plane. The relative directions of these 
${\bm S_{r}}$'s are fixed in opposite, but the direction of the pair is not strongly fixed in 
the $ab$ plane.
The other ${\bm S_{r}}$ illustrated by broken red arrow is directed along the $c$ axis.

\begin{figure}[t]
		\begin{center}
			\includegraphics[width=7cm]{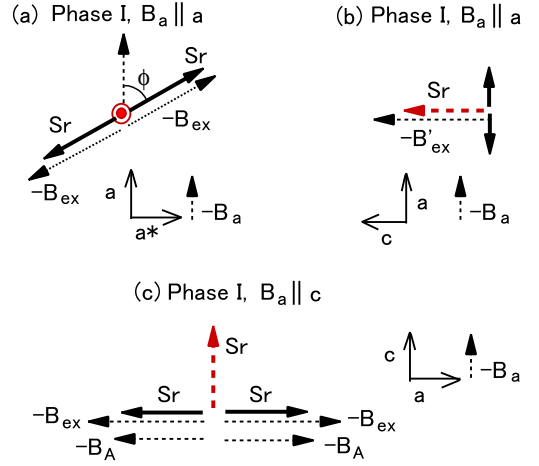}
		\end{center}
		\caption{(Color online) The T-structure spins under 
		a weak applied external flux density ${\bm B_{a}} \parallel a$, 
		projected parallel to (a) $c$ 
		axis and (b) $a^{*}$ axis. Here, ${\bm B_{ex}}$ and ${\bm B'_{ex}}$ is the internal effective 
		flux density originating from exchange interactions which 
		acts on ${\bm S_{r}}$ at each trimer. (c) The T-structure spins projected parallel to $ab$ plane
		 under the ${\bm B_{a}} \parallel c$. ${\bm B_{A}}$ is the effective anisotropic flux density. 
		 The minus signs of the flux densities denote that the magnetic moments and the spins are in
		 opposite direction.
		}
		\label{Spin_Sus}
		\end{figure}

Let us estimate the effective exchange flux densities and anisotropic field
from $\chi_{a}$ (${\bm B} \parallel a$) and
$\chi_{c}$ (${\bm B} \parallel c$).
A set of T-structured ${\bm S_{r}}$'s
under the very weak applied flux density ${\bm B}_{a} \parallel a$ is depicted in 
Fig.~\ref{Spin_Sus} (a).
In this figure, ${\bm B_{ex}}$'s indicate the effective flux densities which act on the pair 
of ${\bm S_{r}}$'s.
The angle $\phi$ is the angle between the pair and $a$ axis.
Figure~\ref{Spin_Sus} (b) displays the same T-structure ${\bm S_{r}}$'s
projected parallel to $a^{*}$ axis. In this figure, ${\bm B}_{ex}'$ denotes the effective flux 
density which acts on the ${\bm S_{r}}$ depicted by the broken red arrows.
Because the in-plane anisotropy is weak, $\phi$ would be equally distributed
over the range from $-\pi/2$ to $\pi/2$ due to domain structure.
The magnetic susceptibility arising from the pairs become to be a mixture of longitudinal 
susceptibility and transverse susceptibility when the ${\bm B_{a}}$ is applied along the
$a$ axis. 
The $\chi_{a}$ expected is
	\begin{align}
		\chi_{a}(0{\rm K})&=2\times \frac{1}{9}N_{\rm A}(2\mu_{\rm B}S_{r})B_{ex}^{-1\,}\frac{1}{\pi}
			\int_{-\pi/2}^{\pi/2}\sin^{2}\phi\,d\phi \nonumber \\
				&\quad+\frac{1}{9}N_{\rm A}(2\mu_{\rm B}S_{r})B_{ex}'^{-1}\notag
	\end{align} 
in a unit of J/(T$^{2}$ Gd-mol).  
Here, $N_{\rm A}$ is the Avogadro number. 
As we mentioned later, the ${\bm B}_{ex}$ and
${\bm B}_{ex}'$ are induced at $T_{2}$ and $T_{1}$, respectively.
Since the $T_{1}$ and $T_{2}$ are approximately equal,
$B_{ex}$ would be approximately equal to $B_{ex}'$. Therefore, we assume
 $B_{ex}=B_{ex}'$.
Thus,
	\begin{align}
		\chi_{a}(0{\rm K})=\frac{2}{9}N_{\rm A}(2\mu_{\rm B}S_{r})B_{ex}^{-1}.
		\label{Kaiperpen}	
	\end{align} 
The $\chi_{a}(1.8{\rm K})$ observed is 1.25 emu/(Gd-mol) as shown in Fig.~\ref{Sus_fit}.
This is converted into 2.24 $\mu_{\rm B}{\rm T}^{-1}$.
Therefore, the effective field $B_{ex}$ is estimated to be 2.08 T.
On the other hand, when the field is applied along the $c$ axis, as shown in Fig.~\ref{Spin_Sus} (c),
the ${\bm S_{r}}$ directed along the $c$ axis does not contribute to magnetic susceptibility at 0 K,
and only the pair of ${\bm S_{r}}$'s directed in the $ab$ plane contribute to the susceptibility, being 
affected by ${\bm B_{A}}$'s.
In this case, $\chi_{c}(0{\rm K})$ would be approximately given by,
	\begin{align}
 		\chi_{c}(0{\rm K})=2\times \frac{1}{9}N(2\mu_{\rm B}S_{r})
 		\left(B_{ex}+B_{A}\right)^{-1}.
 		\label{Kaipara}	 
	\end{align}  
The actual $\chi_{c}(1.8 {\rm K})$ observed is 0.700 emu/(Gd-mol) as shown in Fig.~\ref{Sus_fit}.
This is converted into 1.25 $\mu_{\rm B}{\rm T}^{-1}$.
Substituting this and $B_{ex}=2.08$ T into Eq.~\ref{Kaipara}, $B_{A}$ is obtained to be 1.64 T.
The gain in the anisotropic energy for ${\bm S_{r}}$'s which directed in the $ab$ plane is 
$k_{\rm B}^{-1}(2\mu_{\rm B}S_{r})B_{A}=23.1$ K per ${\bm S_{r}}$.
This is 8.5 times larger than that estimated from electromagnetic interaction before.
If we assume the 120$^{\circ}$ structure parallel to the $ab$ plane, the ratio
$\chi_{a}(0{\rm K})/\chi_{c}(0{\rm K})$ is expected to be 0.89 considering $B_{A}$.
This shows significant disagreement with the ratio 1.79 experimentally obtained at 1.8 K.

Assuming the T-structure, we have estimated $B_{ex}$ and $B_{A}$ from the low temperature 
limits of $\chi$'s. 
We would be able to calculate $E_{g}$ in Eq.~\ref{Gapenergy} from these.
Considering that the number of ${\bm S_{r}}$'s is $2/9$ moles, the energy 
$(nk_{\rm B})^{-1}\hbar\omega_{A}=(nk_{\rm B})^{-1}2\mu_{\rm B}B_{A}$ is obtained to be 9.90 K. 
If we assume that $B_{ex}$ is determined only by the exchange interactions from the nearest 
neighbor ${\bm S_{r}}$'s, 
$(nk_{\rm B})^{-1}\hbar\omega_{ex}=(nk_{\rm B})^{-1}(2\times2\mu_{\rm B}B_{ex})=25.2$ K. 
Substituting these into Eq.~\ref{Gapenergy}, we obtain $E_{g}=24$ K. This agrees with that 
obtained from the temperature dependence of $\chi$ before. 

\subsection{Spin structure in phase II}

If we consider only the interactions among the three ${\bm S_{r}}$'s and easy plane anisotropy,
the Hamiltonian is written as
	\begin{align}
		\mathscr{H}=J\sum_{ij}{\bm S_{i}}{\bm S_{j}}
						-D\sum_{i=1,2,3}(S_{i}^{z})^{2},\quad(D<0).
						\label{EQ:T_struct}
	\end{align}
Here, the first summation runs over $ij=12, 23, 31$.
Equation~\ref{EQ:T_struct} shows that the T-structure has two types of independent operations,
which give degeneracies in energy.
One is the operations with respect to the 2D rotation of the pair of ${\bm S_{r}}$'s 
indicated by solid black arrows in Fig.~\ref{T_Structure} around the $c$ axis, 
and the other is the conversion operation of the directions of the ${\bm S_{r}}$
depicted by the broken red arrows with respect to the symmetry plane $ab$.
The former type form a 2D rotational group ${\bm S_{1}}$, and the latter type
forms a cyclic group of order two ${\bm Z_{2}}$ with the identity operator.
This suggests that these two kinds of degeneracies lead to the successive phase 
transitions.

\begin{figure}[h]
		\begin{center}
			\includegraphics[width=5cm]{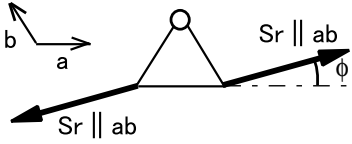}
		\end{center}
		\caption{The spin structure in phase II on the triangle of the trimers. 
		A pair of ${\bm S_{r}}$'s in opposite 
		directions is directed parallel to the $ab$ plane, and the open circle denotes the partial 
		disordered site (trimer). The angle $\phi$ rotation around the $c$ axis is an element of the 
		${\bm S_{1}}$ group (see text).
		}
		\label{Spin_Phase2}
		\end{figure}

We suggest phase II is the phase wherein only ${\bm S_{1}}$
symmetry is broken, as shown in Fig.~\ref{Spin_Phase2}. 
In this figure, a collinear pair of ${\bm S_{r}}$'s in the opposite directions is directed
in the $ab$ plane and the angle $\phi$ is fixed in the Gd--Al layer. 
The open circle in Fig.~\ref{Spin_Phase2} denotes the partial disorder site
(trimer).
Since the anisotropy in the $ab$ plane is small, the directions of the pair may be distributed 
in the $ab$ plane by the domain structure at low fields. 
However, when the fields increase by certain degree, the directions of the pairs would be
oriented in the direction perpendicular to the applied field, or in the easy direction 
to magnetize. 
Then the pair would show transverse magnetic susceptibility. 
Actually, as shown in Figs.~\ref{MTandCT} (b) and (c), the magnetization of 
Gd$_{3}$Ru$_{4}$Al$_{12}$ under the field shows weak temperature dependence in phase II, 
not being dependent on the directions of applied fields.
This is a feature of transverse magnetic susceptibility. 
When temperature becomes lower than $T_{1}$, ${\bm Z_{2}}$ degeneracy
is lifted and the spin structure changes into the T-structure. In association with this change,
the component of the longitudinal magnetic susceptibility would be added to $\chi_{c}$.
Actually, the magnetization at 0.5 and 1 T in Fig.~\ref{MTandCT} (c) exhibits rapid decrease with
decreasing temperature below $T_{1}$. This is considered to be the contribution of 
longitudinal magnetic susceptibility.
As we mentioned before, phase II does not appear at a lower temperature side of phase I,
while phase I appears at a lower temperature side of phase II (Fig.~\ref{PhaseD}). 
This is easily understood if we assume the above partial disorder in phase II.

\begin{figure}[h]
		\begin{center}
			\includegraphics[width=7cm]{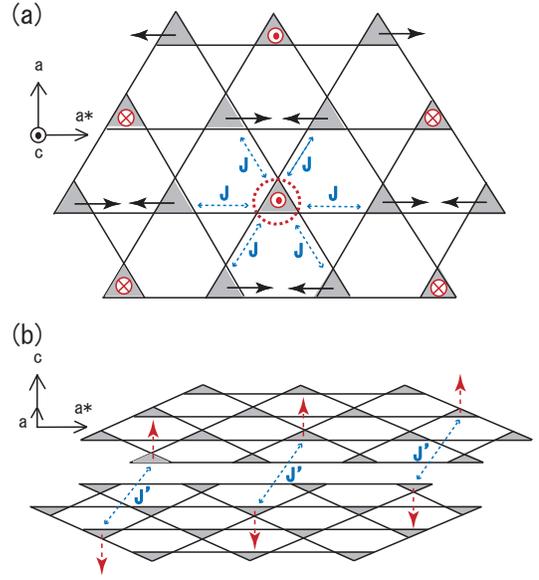}
		\end{center}
		\caption{(Color online) A possible ${\bm S_{r}}$ structure in phase I. The gray
		triangles denote the trimers. In this figure, spiral modulations are not considered.
		(a) The ${\bm S_{r}}$'s illustrated by the solid black arrows are directed in the $ab$ plane
		 and those by the red $\odot$ are in the opposite $c$ axis direction and 
		 the others by $\otimes$ are in the negative $c$ axis directions. 
		$J$ denotes the exchange integral between nearest 
		${\bm S_{r}}$'s in the same Gd--Al layer. 
		(b) The broken red arrows indicate ${\bm S_{r}}$'s directed along the $c$ axis. 
		$J'$ denotes the exchange integral between the nearest ${\bm S_{r}}$'s
		on the nearest neighbor Gd--Al layer. In this panel, only a part of the ${\bm S_{r}}$'s
		is illustrated for easy look.
		}
		\label{Spin_struct}
		\end{figure}

So far the spin structure of Gd$_{3}$Ru$_{4}$Al$_{12}$ has not been 
determined by microscopic measurements. However, we discuss a possible orientation
to examine the consistency between the ${\bm S_{r}}$ structures shown in Figs.~\ref{T_Structure} 
and \ref{Spin_Phase2} and the successive phase transitions mentioned above.
We illustrate the ${\bm S_{r}}$ orientation in phase I on a Gd--Al layer in 
Fig.~\ref{Spin_struct} (a). The small gray triangles indicate trimers. 
The black arrows denote the ${\bm S_{r}}$'s directed in the $ab$ plane, and the red 
$\odot$ and $\otimes$ indicate ${\bm S_{r}}$'s directed along the $c$ axis. 
As shown in Fig.~\ref{Spin_struct} (a), each triangle of the trimers exhibits a T-structure. 
Let us note of the ${\bm S_{r}}$ surrounded by the broken red circle 
in Fig.~\ref{Spin_struct} (a). 
This ${\bm S_{r}}$ receives exchange interactions $J$ from six nearest 
neighbor ${\bm S_{r}}$'s in the same layer, but these exchange interactions are canceled out
with each other.
Such condition would lead to a partial disorder in phase II, as illustrated in Fig.~\ref{Spin_Phase2}. 
Figure \ref{Spin_struct} (b) shows the ${\bm S_{r}}$'s on two nearest neighbor
Gd--Al layers. The broken arrows denote the AFM interlayer exchange integral $J'$ 
which acts between
the nearest ${\bm S_{r}}$'s on the nearest layers. This interaction generates spontaneous
${\bm S_{r}}$'s at the partially disordered sites below $T_{1}$.

\begin{figure}[h]
		\begin{center}
			\includegraphics[width=7.5cm]{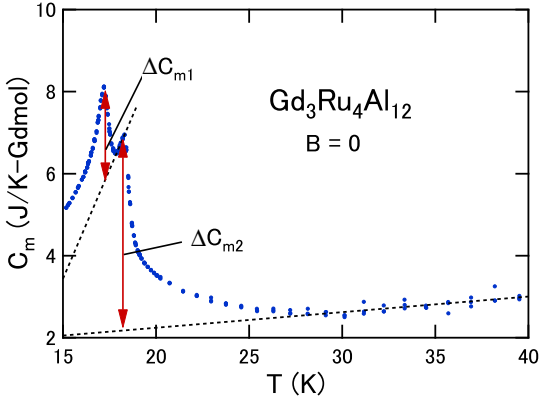}
		\end{center}
		\caption{(Color online) Magnetic specific heat jumps of Gd$_{3}$Ru$_{4}$Al$_{12}$ at zero field.
		Data are taken from the reference
		\cite{Nakamura2018}.
		The dotted lines are fits to lines in the ranges 17.9--18.3 K and 28.2--39.5 K.		 			$\Delta C_{m1}$ and $\Delta C_{m2}$ are obtained as
		2.35 and 4.78 J/(K Gd-mol), respectively.
		}
		\label{SpecHeatJump}
		\end{figure}

As shown in Fig.~\ref{Spin_struct}, the number of ${\bm S_{r}}$'s that order at $T_{2}$
is expected to be two times larger than the number of ${\bm S_{r}}$'s that order at $T_{1}$.
According to the mean field theory of second order phase transitions, the jumps of the magnetic
specific heat $\Delta C_{m}$ at $T_{1}$ and at $T_{2}$ are expected to be 
proportional to the numbers of ${\bm S_{r}}$'s, which order at each temperature. 
We present the magnetic specific heat $C_{m}$ of Gd$_{3}$Ru$_{4}$Al$_{12}$ at zero field in the 
vicinity of phase transition temperatures in Fig.~\ref{SpecHeatJump}. The dotted lines in this 
figure are fits to lines. 
The jumps $\Delta C_{m1}$ at $T_{1}$ and $\Delta C_{m2}$ at $T_{2}$ are found to be 2.35
and 4.78 J/(K Gd-mol), respectively, or 0.282$R$ and 0.574$R$ in the unit of gas constant
$R$, respectively. 
The ratio $\Delta C_{m2}/\Delta C_{m1}$ obtained is 2.03,
which agrees well with that expected from Fig.~\ref{Spin_struct}.

\subsection{Spin structure in phase III and the anisotropic energy}

\begin{figure}[t]
		\begin{center}
			\includegraphics[width=6cm]{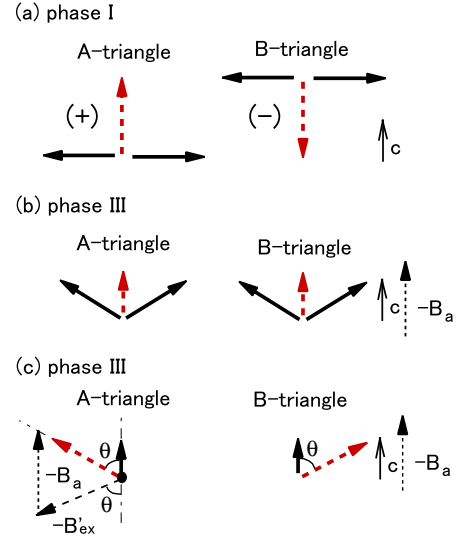}
		\end{center}
		\caption{(Color online) (a) The T-structures of ${\bm S_{r}}$'s in phase I at zero field. 
		The resultant spins ${\bm S_{r}}$'s depicted by the broken red arrows on triangles are directed 
		along the $c$ axis and interact with each other antiferromagnetically.
		The signs ($+$) and ($-$) correspond to the degrees of freedom of ${\bm Z_{2}}$ degeneracy. 
		(b) The canted T-structures in phase III under an applied external flux density 
		${\bm B_{a}}\parallel c$. 
		The broken red arrows indicate the occurrence of spin flopping. The minus sign of ${\bm B_{a}}$
		denotes the opposite directions of the ${\bm S_{r}}$'s and the magnetic moments.
		(c) The ${\bm S_{r}}$'s in panel (b) projected to the direction perpendicular 
		to (b) and parallel to the $ab$ plane.
		}
		\label{PhaseIII}
		\end{figure}

As shown in Fig.~\ref{PhaseD}, we have observed the additional phase III in the intermediate
field range when fields are directed along the $c$ axis. 
The hysteresis loops shown in Fig.~\ref{MH5K_C} (c) indicates that the phase I/phase III
transition is first order.  
We present the change in the spin structures assumed in association with this transition 
in Fig.~\ref{PhaseIII} (c).
In this figure, panel (a) denotes the T-structures on A-triangle and B-triangle. These two
triangles are on the nearest neighbor layers as shown in Fig.~\ref{Spin_struct} (b).
In the absence of the field, the ${\bm S_{r}}$'s denoted by the red broken arrows on each triangle 
are directed along the $c$ axis and canceled out with each other.
Between these two ${\bm S_{r}}$'s the AFM interaction $J'$ is acting (Fig.~\ref{Spin_struct}). 
When the external flux densities ${\bm B_{a}}$ are applied along the $c$ axis as illustrated in 
Fig.~\ref{PhaseIII} (b), the ${\bm S_{r}}$'s depicted by the broken red
arrows occur to be spin flopping and phase III appears.
Figure \ref{PhaseIII} (c) shows the ${\bm S_{r}}$'s in Fig. \ref{PhaseIII} (b)
projected in a direction perpendicular to Fig. \ref{PhaseIII}(b) and parallel to 
the $ab$ plane. With further increasing the field, the AFM coupling between the broken red arrows in 
Fig.~\ref{PhaseIII} is broken and the phase III/phase II transition occurs.

\begin{figure}[h]
		\begin{center}
			\includegraphics[width=6cm]{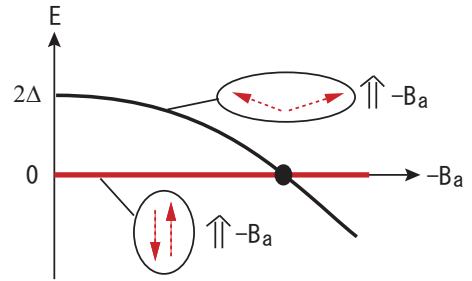}
		\end{center}
		\caption{(Color online) (a) Field dependence of the energy of 
		the pair of ${\bm S_{r}}$'s. The red bold line denotes the change in the energy
		when the pair is assumed to be directed along the $c$ axis. The solid black curve indicates the
		energy when the pair is assumed to be parallel to the $ab$ plane at zero field and canted 
		by the applied flux density.
		}
		\label{Spinflop}
		\end{figure}

The spin flopping illustrated in Figs.~\ref{PhaseIII} (a) and (b) occurs at 1.25 T as evident in
Fig.~\ref{PhaseD} (c). 
We define the angle $\theta$ as shown in Fig.~\ref{PhaseIII} (c), and assume that
the anisotropic energy acts on the ${\bm S_{r}}$'s depicted by the red broken arrows as
$\Delta_{Sr} (1-\cos^{2}{\theta})$ in a unit of J per ${\bm S_{r}}$.
Figure \ref{Spinflop} represents the field dependence of the energy of the pair.
When the pair is assumed to be directed in the $c$ axis, the energy of the pair is field independent.
On the other hand, when the pair is assumed to be directed in the $ab$ plane at zero field,
the magnetization $2\mu_{\rm B}S_{r}(B_{a}/B_{ex}')$ is induced by ${\bm B_{a}}$.
Then the energy of the pair is approximately written as,
		\begin{align} 
			E(B_{a}) = 2\Delta_{Sr} - \mu_{\rm B} S_{r}  (B_{a}^{2}/B_{ex}'), 
			\notag
		\end{align}
in the weak field range. Since spin flopping occurs at $E=0$,
$\Delta_{Sr}$ is given by
		\begin{align} 
			\Delta_{Sr} = (1/2)\mu_{\rm B} S_{r}  (B_{t}^{2}/B_{ex}'). \notag
			\label{Anisoenrgy}
		\end{align}
Here, the transition field is $B_{t}=1.25$ T and $B'_{ex}$($\doteqdot B_{ex}$) is  2.08 T,
as we mentioned before.
Then, the anisotropic energy $\Delta_{Sr}=3.7\times 10^{-23}$ J per ${\bm S_{r}}$, or
2.6 K per ${\bm S_{r}}$ is obtained.
In phase III, the pair of ${\bm S_{r}}$'s depicted by broken
red arrows in Fig~\ref{PhaseIII} (b) and (c) are approximately oriented along the 
high energy directions concerning the anisotropic energy in phase III.
Therefore, these ${\bm S_{r}}$'s tend to eliminate AFM coupling and change their directions along 
the $c$ axis in the high field range due to the anisotropic energy.
Thus, the phase III/phase II boundary shifts to a lower field side.
As evident in Fig.~\ref{MH5K_C} (b), hysteresis loops are observed in magnetization 
at the phase III/phase II transition points. Therefore, this transition is first order.
On the other hand, the anisotropic flux density ${\bm B_{A}}$ stabilizes AFM phase II when the
applied fields are directed along the $c$ axis, and it would shift the phase II/PM phase
boundary to a higher field side.
The anisotropic flux density $B_{A}$ is obtained as 1.64 T. This approximately
agrees with the shift of phase II/PM phase boundary at 1.8 K as evident in Fig.~\ref{PhaseD}. 
These are the reasons why phase II occupies the wide region of the phase diagram
for ${\bm B}\parallel c$.
As shown in the inset of Fig.~\ref{Sus_fit}, $\chi_{c}$ deviates
from the $T^{2}$ behavior below 4.5 K. This deviation may arise from $\Delta_{Sr}$. 
This energy is sufficiently low compared to $T_{1}$, but it can affect the magnetic
susceptibility in the approximate range $T\lesssim 2\Delta_{Sr}$.

\subsection{Low energy magnetic excitations and long period structures}

		\begin{figure}[h]
		\begin{center}
			\includegraphics[width=8cm]{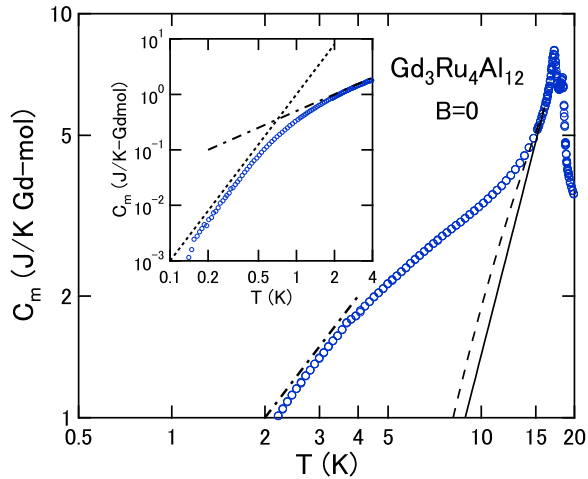}
		\end{center}
		\caption{(Color online) Magnetic specific heat $C_{m}$ of Gd$_{3}$Ru$_{4}$Al$_{12}$
		at zero field on a log$T$-log$C_{m}$ plot. Experimental data (open blue circles) are 
		taken from the reference
		\cite{Nakamura2018}.
		The solid line is the eye guide, which indicates the slope of $C_{m}\propto T^{3}$.
		The broken line is the eye guide, which denotes $C_{m}$ with $E_{g}=24$ K (see text).
		The dotted-broken line is the eye guide, which indicates the slope of $C_{m}\propto T$.
		The inset shows $C_{m}$ in the low temperature range on a log$T$-log$C_{m}$ plot.
		The dotted line and dotted-broken line in the inset are eye guides, which indicate 
		the slopes of $C_{m}\propto T^{3}$ and $C_{m}\propto T$, respectively. 
		}
		\label{LowT_C}
		\end{figure}

The magnetic susceptibility of Gd$_{3}$Ru$_{4}$Al$_{12}$ in phase I can be explained by the 
three-dimensionally propagating spin waves or magnons. On the other hand, the specific heat of Gd$_{3}$Ru$_{4}$Al$_{12}$ in phase I shows peculiar behaviors. Figure~\ref{LowT_C} 
displays the magnetic specific heat
$C_{m}$ of Gd$_{3}$Ru$_{4}$Al$_{12}$ at zero field on a log$T$-log$C_{m}$ plot.
In this figure, the open blue circles indicate experimental data of $C_{m}$.
It is well known that 3D AFM magnons contribute to the specific heat in proportion
to $T^{3}$ when the magnon dispersion is gapless
\cite{QTS}. 
The solid line in Fig.~\ref{LowT_C} is the temperature dependence of $C_{m}$ expected from 
AFM magnons without energy gap.
When magnon dispersion is written by Eq.~\ref{Dispersion}, $C_{m}$ is given by,
		\begin{align} 
			C_{m} \propto e^{-(E_{g}/T)} \left[
			\frac{5}{2}\,T^{\frac{3}{2}}+2E_{g}\,T^{\frac{1}{2}}
			+\frac{2}{3} (E_{g})^{2}\,T^{-\frac{1}{2}} \right].
			\notag
		\end{align}
The broken line in Fig.~\ref{LowT_C} is $C_{m}$ of AFM magnons with energy gap $E_{g}=24$ K.
The exponential factor on the right side mainly determines the temperature dependence.
Both calculated data are normalized at 16 K, which is the high temperature end of the fitting range
of magnetic susceptibility shown in Fig.~\ref{Sus_fit}.
As evident in Fig.~\ref{LowT_C}, both contributions from magnons rapidly decrease 
with decreasing temperature,
therefore, we cannot reproduce $C_{m}$ experimentally observed by adding these two at any ratio. 
The actual $C_{m}$ of Gd$_{3}$Ru$_{4}$Al$_{12}$ decreases more slowly with decreasing temperature. 
This means that certain low energy excitations other than magnons exist in phase I at low temperatures.
It is known that a heavy fermion often coexists with AFM magnons
\cite{Furuno1985,Kadowaki1986}. 
However, the low energy excitation in Gd$_{3}$Ru$_{4}$Al$_{12}$ is not a heavy fermion.
The inset in Fig.~\ref{LowT_C} displays $C_{m}$ in the low temperature range
on a log$T$-log$C_{m}$ plot. The dotted-broken line and the dotted line are the eye guides
which indicate the slopes of the functions $C_{m}\propto T$ and $C_{m}\propto T^{3}$, respectively.
The temperature dependence of $C_{m}$ approximately follows $T^{3}$ behavior below 0.5 K, being contradictory to heavy fermion state.
In addition to this, no $T^{2}$ behavior is observed in the low temperature electrical resistivity 
of Gd$_{3}$Ru$_{4}$Al$_{12}$
\cite{Nakamura2018}.
It is probable that certain low energy quasi-particles which do not contribute to magnetization 
may contribute to the low temperature $C_{m}$ of Gd$_{3}$Ru$_{4}$Al$_{12}$.
For example, vortexes proposed by Kawamura and Miyashita may be one of the candidates
of low energy excitation
\cite{KawamuraMiyashita1984}.

In the present paper, we have investigated basic properties and spin structures of 
Gd$_{3}$Ru$_{4}$Al$_{12}$ using macroscopic measurements.
It is inferred that spiral spin structures may be induced by the competition between 
far and near neighbors interactions.
However, such long period structures and detailed of the low energy excitations should be 
investigated by microscopic measurements.  
Unfortunately, Gd ions are good absorbers of neutrons, but investigations
by resonant X-ray diffraction may be applicable.
For example, the cycloidal magnetic structure of GdRu$_{2}$Al$_{10}$ has been determined
by this method
\cite{Matsumura2017}.

\section{Summary}

We grew single crystals of Gd$_{3}$Ru$_{4}$Al$_{12}$ with the distorted kogome lattice 
structure wherein stacked AFMTL is formed in association with spin trimerization 
and the geometrical frustration and interaction-compete-type frustration coexist
via the RKKY interaction.
Gd$_{3}$Ru$_{4}$Al$_{12}$ is found to be a spin system that has a certain degree of strong
easy plane-type anisotropy and interlayer interactions.
It is highly probable that a partial disorder occurs in this ${\bm S_{r}}$'s system.
With decreasing temperature, first, the AFM long-range order, wherein ${\bm S_{r}}$'s 
are directed in the $ab$ plane, occurs at $T_{2}$ and IMT phase II appears.
This phase is a partial disordered phase wherein 1/3 of ${\bm S_{r}}$'s is not arranged
and only ${\bm S_{1}}$ degeneracy is lifted.
With further decreasing temperature, ${\bm S_{r}}$'s at the disordered sites exhibit the AFM
order wherein a part of them are oriented along the $c$ axis at $T_{1}$.
In association with this transition, ${\bm Z_{2}}$ degeneracy is lifted.
Thus, the noncollinear T-structure of ${\bm S_{r}}$ is formed in phase I.
We found an additional phase III in intermediate fields directed along the $c$ axis,
where spin flopping has occurred in the part of ${\bm S_{r}}$'s which is directed in the $c$ axis
at zero field. 
The temperature dependence of the magnetic susceptibilities is well explained
by the contribution of three dimensionally propagating magnons. On the other hand, the specific
heat in this phase is not understandable only as the contribution of magnons. 
Certain magnetic excitations other than magnons or heavy fermion may exist owing to the frustration.

\section*{Acknowledgement}
The authors thank S. Tanno, K. Hosokura, A. Ogata, M. Kikuchi, H. Moriyama and N. Fukiage,
Tohoku University,
for supporting our low-temperature experiments.

\end{document}